\documentclass[pre,twocolumn]{revtex4}
\usepackage{url}
\usepackage{color}
\usepackage{amsmath}
\usepackage{amssymb}
\usepackage{graphicx}
\usepackage{algorithm,algpseudocode}

\begin{document}

\title{Fast algorithm for generating random bit strings and multispin coding for directed percolation}

\author{Hiroshi Watanabe$^1$}
\email{hwatanabe@issp.u-tokyo.ac.jp}
\thanks{Corresponding author}

\author{Satoshi Morita$^1$}
\author{Synge Todo$^{2,1}$}
\author{Naoki Kawashima$^1$}

\affiliation{%
$^1$ The Institute for Solid State Physics, The University of Tokyo,
Kashiwanoha 5--1--5, Kashiwa, Chiba 277--8581, Japan
}

\affiliation{%
$^2$ Department of Physics, The University of Tokyo, Tokyo 113--0033, Japan
}

\begin{abstract}
We present efficient algorithms to generate a bit string in which each bit is set with arbitrary probability. By adopting a hybrid algorithm, i.e., a finite-bit density approximation with correction techniques, we achieve 3.8 times faster random bit generation than the simple algorithm for the 32-bit case and 6.8 times faster for the 64-bit case. Employing the developed algorithm, we apply the multispin coding technique to one-dimensional bond-directed percolation. The simulations are accelerated by up to a factor of 14 compared with an optimized scalar implementation.
The random bit string generation algorithm proposed here is applicable to general Monte Carlo methods.
\end{abstract}

\maketitle

\section{Introduction}
A computer performs computation in the unit of words, which consist of a fixed number of bits.
Multispin coding (MSC) is a technique to reduce the memory space and execution time in Monte Carlo simulations by packing information of spins into a computer word. 
In modern CPUs, the width of a computer word is 32 or 64, and bit operations between words are usually executed in one cycle. Therefore, a simulation of a system with discrete degrees of freedom can be markedly accelerated by effectively utilizing MSC and bit operations. The MSC technique was first mentioned by Friedberg and Cameron in 1970~\cite{Friedberg1970}, but Rebbi and co-workers proposed MSC in the form widely used today~\cite{Creutz1979, Rebbi1980}. Since then, the MSC of spin systems, especially for the Ising model, has been extensively studied, mainly in the 1980s~\cite{Zorn1981, Wansleben1984, Bhanot1986, Wansleben1987, Kikuchi1987, Ito1988, Kawashima1993}. A special-purpose computer for Ising models was developed on the basis of MSC~\cite{Taiji1988}. The MSC technique has also been applied to a cellular automaton~\cite{Silva1988}.
The MSC is naturally applied to a system which has Ising-like or finite degrees of freedom.
One of such systems is the directed percolation (DP),
which is the directed version of isotropic percolation.
Consider a lattice consisting of sites and bonds.
Each bond is open with some probability and blocked otherwise.
In isotropic percolation, a site becomes active when it is connected to
another active site with an open bond. In DP, however, a site can be activated
only along a preferred direction.
The universality class of DP is different from that of the isotropic percolation~\cite{Jensen1999}.
The transition from a laminar flow to turbulence is expected to belong to the DP universality class,
which has been confirmed in several experiments~\cite{Takeuchi2007, Sano2016}.
The sites in DP have two states, active and inactive.
Therefore, it is natural to consider applying the MSC technique to DP simulation.
The MSC of DP was reported by Hinrichsen and {\'{O}}dor in 1999~\cite{Hinrichsen1999}.
They adopt the replica-parallel method, \textit{i.e.}, the $N_\mathrm{bit}$ independent replicas are encoded in one integer, where $N_\mathrm{bit}$ is the bit-length of the integer of the computer.
By using the same random numbers on each replica, they updated $N_\mathrm{bit}$ replicas simultaneously.
The special-purpose hardware for generating random numbers was also developed and it was also applied to DP~\cite{Rohrbach1997, Odor1999}.

In this manuscript, we present efficient algorithms to generate a bit string in which each bit is set with arbitrary probability. These algorithms allow us to implement MSC for one-dimensional bond-directed percolation on the basis of the vector-parallelism. The vector-parallel MSC is the non-trivial parallelism where $N_\mathrm{bit}$ sites in one system are encoded in one integer. The MSC technique accelerates the simulations by up to a factor of 14 compared with an optimized scalar implementation. The rest of the article is organized as follows. The random bit string generation algorithm is described in the next section. The application of the algorithm to 1d-BDP with the MSC technique is described in Sec.~\ref{sec:dp}. Section~\ref{sec:summary} is devoted to a summary and discussion. The associated code is available at~\cite{github}.

\section{Random bit string generation algorithm}

We consider algorithms to generate a bit string with length $N_\mathrm{bit}$ such that each bit is $1$ with a given probability $p$ and otherwise $0$.
A simple algorithm is shown in Algorithm~\ref{alg:simple}.
Here, $U_r(a,b)$ is a stochastic real variable distributed uniformly from $a$ to $b$.
The logical left shift is denoted by $\ll$. 
The operation $\ll k$ results in a bit shift by $k$ bits, which is multiplication by $2^k$.
Excess bits shifted off to the left are discarded.
The operation $a \lor b$ denotes the bitwise OR between bit strings $a$ and $b$.
This simple algorithm involves generating random numbers $N_\mathrm{bit}$ times regardless of the value of $p$, and therefore, it is expensive to use it for the MSC technique.
In order to reduce the computational cost to generate a random bit string, we propose two approaches.
The first approach is effective when the number of set bits in the bit string is small.
We propose two such algorithms, one is the binomial-shuffle algorithm described next section
and the other is the Poisson-OR algorithm which is described in Sec.~\ref{sec:po}.
The second approach is to utilize the fact that the bit string generated by the standard pseudo-random generator contains set bits with probability 0.5.
This algorithm, which is described in Sec.~\ref{sec:fd}, is effective when the number of digits in binary notation of a probability $p$ is small.
In Sec.~\ref{sec:correction}, we propose a hybrid algorithm of the two approaches.

\begin{algorithm}
\begin{algorithmic}[1]
\State $s \leftarrow 0$
\For{ $i=0$ to $N_\mathrm{bit}-1$}
\If {$U_r(0,1) < p$}
\State $s \leftarrow  s \lor (1 \ll i)$
\EndIf
\EndFor
\end{algorithmic}
\caption{Simple Algorithm}
\label{alg:simple}
\end{algorithm}

\subsection{Binomial-Shuffle Algorithm}

In the simple algorithm, it is necessary to generate $N_\mathrm{bit}$ random numbers
regardless of the probability $p$.
Since the number of set bits is about $p N_\mathrm{bit}$, we can reduce the number of random numbers generated by first determining the number of set bits and then shuffling their positions.
We refer to this algorithm as the binomial-shuffle algorithm.
First, we determine the number of set bits $m$ out of $N_\mathrm{bit}$ with probability $p$.
This is a random number following the binomial distribution of $N_\mathrm{bit}$ trials with probability of success $p$.
After determining the number of set bits, we choose their positions.
Adopting Floyd's sampling algorithm, this selection process involves random number generation $m$ times. 
The pseudocode of the binomial-shuffle algorithm is shown in Algorithm~\ref{alg:shuffle}.
The operation $a \land b$ denotes the bitwise AND.
Here, $\mathrm{B}(n,p)$ is a stochastic integer variable following the binomial distribution with parameters $n$ and $p$, where $n$ is the number of trials and $p$ is the probability of success.
We can implement the function $\mathrm{B}(n,p)$ with $O(1)$ complexity by adopting
Walker's method of aliases~\cite{Walker1977,Knuth1997}.
$U_d(a,b)$ is a stochastic integer variable distributed uniformly from $a$ to $b$.
Since we need one random number to determine the number of set bits
and $p N_\mathrm{bit}$ random numbers to shuffle the positions of the bits on average, this algorithm involves
the generation of $p N_\mathrm{bit} + 1$ random numbers.

\begin{algorithm}
\begin{algorithmic}[1]
\State $s \leftarrow 0$
\State $m \leftarrow \mathrm{B}(N_\mathrm{bit},p)$
\For{ $i=N_\mathrm{bit}-m$ to $N_\mathrm{bit}-1$}
\State $k \leftarrow U_d(0,i)$
\State $t \leftarrow 1 \ll k$
\If {$(s \land t) \neq 0$}
\State $s \leftarrow s \lor (1 \ll i)$
\Else
\State $s \leftarrow s \lor t$
\EndIf
\EndFor
\end{algorithmic}
\caption{Binomial-Shuffle Algorithm}
\label{alg:shuffle}
\end{algorithm} 

\subsection{Poisson-OR Algorithm} \label{sec:po}

The binomial-shuffle algorithm reduces the cost of generating a bit string by first determining the number of set bits in $O(1)$ complexity and next determining their position in $O(p N_\mathrm{bit}$) complexity.
Todo and Suwa proposed an efficient algorithm to generate a bit string in which each bit is set with some probability $p_k$, where $k$ is the index of the bits~\cite{Todo2013}. 
They employed the space-time interchange technique together with Walker's method of aliases and achieved the generation of such a bit string in $O(\sum_k p_k)$ complexity. This algorithm was applied to spin systems with long-range interactions and achieved $O(N)$ complexity for the cluster Monte Carlo method without introducing any cutoff~\cite{Fukui2009}.
We need a bit string in which all bits are set independently with identical probability.
Since this is a special case of Todo and Suwa's case, we can apply their algorithm to our problem as follows.
Consider an $N_\mathrm{bit}$-length bit string with one of the bits is set randomly.
We generate such bit strings $k$ times and take the bitwise OR between them.
Then each bit of the resulting bit string is $1$ with probability $1 - (N_\mathrm{bit}-1)^k/ N_\mathrm{bit}^k$. We choose the number $k$ following the Poisson distribution with parameter $\lambda$. 
Since the probability that the number of events is $k$ in the Poisson process with parameter $\lambda$
is $\lambda^k \mathrm{e}^{-\lambda}/k!$, 
the probability that each bit in the resulting bit string is set is given by
\begin{equation}
\begin{split}
p &= \sum_{k=0}^{\infty} \left[ 1- \left(\frac{N_\mathrm{bit} - 1}{N_\mathrm{bit}} \right)^k\right] \frac{\lambda^k \mathrm{e}^{-\lambda}}{k!},\\
&= \sum_{k=0}^{\infty} \frac{\lambda^k \mathrm{e}^{-\lambda}}{k!} - 
\sum_{k=0}^{\infty} \left(\frac{N_\mathrm{bit}-1}{N_\mathrm{bit}} \right)^k \frac{\lambda^k \mathrm{e}^{-\lambda}}{ k!},\\
&= 1 - \exp\left(-\lambda/N_\mathrm{bit}\right).
\end{split}
\end{equation}
Therefore, when we choose $\lambda=-N_\text{bit}\log(1-p)$ then each
bit in the resulting bit string is set with probability $p$.
Moreover, each of the bits is mutually independent, which is proved by
mathematical induction and the fact that the probability that specified
$n$ bits in the resulting bit string are zero is given by
\begin{equation}
 \sum_{k=0}^{\infty}
  \left(\frac{N_\text{bit}-n}{N_\text{bit}}\right)^k
  \frac{\lambda^k e^{-\lambda}}{k!}
  = (1-p)^n.
\end{equation}
A more sophisticated proof using the Stirling numbers of the second kind
is given in Appendix.

Then we obtain the following algorithm: 1) determine an integer $k$ following
the Poisson distribution with parameter $\lambda = - N_\mathrm{bit} \log(1-p)$,
2) generate $k$ bit strings in which one of the $N_\mathrm{bit}$ bits is set randomly, and
3) take the bitwise OR between them.
We refer to this algorithm as the Poisson-OR algorithm, whose pseudocode is shown in Algorithm~\ref{alg:or}.
Here, $\mathrm{Poisson}(\lambda)$ is a stochastic integer variable following the Poisson distribution
with parameter $\lambda$. We can generate such an integer with $O(1)$ complexity by adopting Walker's algorithm. We need one random number to determine $k$ and $- N_\mathrm{bit} \log(1-p)$ random numbers
to generate the bit strings for disjunction, and therefore, the Poisson-OR algorithm
involves the generation of $- N_\mathrm{bit} \log(1-p)+1$ random numbers.
While the number of random numbers generated in the Poisson-OR algorithm
is always larger than that in the binomial-shuffle algorithm since $p N_\mathrm{bit} \leq N_\mathrm{bit} \log(1-p)$, they are close to each other when $p$ is small.
Additionally, the loop body of the Poisson-OR algorithm is simpler than that of the binomial-shuffle algorithm. Therefore the Poisson-OR algorithm can be faster than the binomial-shuffle algorithm.

\begin{algorithm}
\begin{algorithmic}[1]
\State $s \leftarrow 0$
\State $k \leftarrow \mathrm{Poisson}(- N_\mathrm{bit} \log(1-p))$
\For{ $i=1$ to $k$}
\State $j \leftarrow U_d(0,N_\mathrm{bit}-1)$
\State $s \leftarrow s \lor (1 \ll j)$
\EndFor
\end{algorithmic}
\caption{Poisson-OR Algorithm}
\label{alg:or}
\end{algorithm} 

\subsection{Finite-Digit Algorithm} \label{sec:fd}

Suppose there are two bit strings $s_1$ and $s_2$ in which each bit is set with probability $p_1$ and $p_2$.
If we make a new random bit string $y = s_1 \land s_2$, then $y$ is a bit string in which each bit is set with a probability $p= p_1 p_2$. Similarly, we can have a bit string $y = s_1 \lor s_2$ in which each bit is set with probability $p = 1 - (1-p_1) (1 -p_2)$. By taking the bitwise AND or the bitwise OR of two random bit strings, we can generate a new bit string in which each bit is set with a new probability.
Consider two random numbers which are generated by the standard random number generator, such as \verb|std::mt19937|. The random numbers can be regarded as random bit strings in which each bit is set with a probability $p=0.5$. Then we can generate a random bit string with a probability $p=0.25$ by taking the bitwise AND of them. We can also generate a random bit string with a probability $p=0.75$ by taking the bitwise OR. In this manner, we can generate a random bit string with a probability $\tilde{p}_n$ by using $n$ random numbers, where $\tilde{p}_n$ is an $n$-digit number in the binary notation.

\begin{figure}[ht]
\includegraphics[width=7cm]{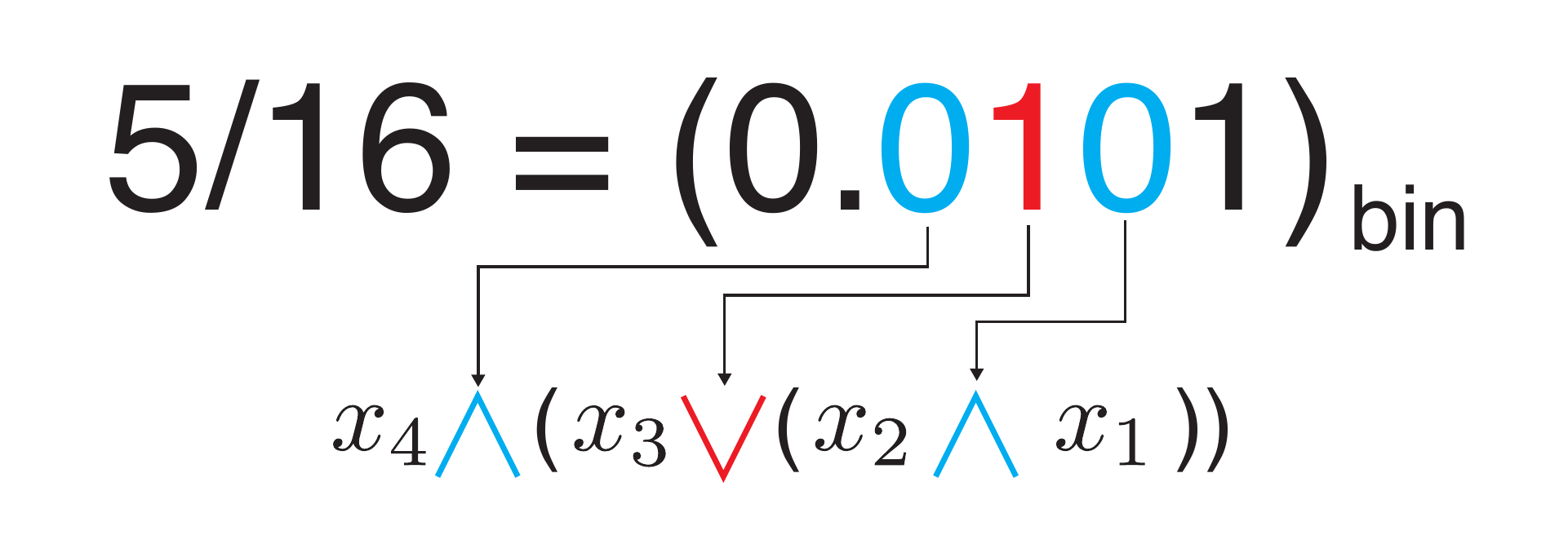}
\caption{
(Color online) Binary notation of the probability and bitwise operations.
Suppose we generate a random bit string in which each bit is set with a probability $\tilde{p}_4 = 5/16 = (0.0101)_\mathrm{bin}$.
Since this probability is expressed by four digits in the binary notation,
we can generate a corresponding bit string by four random numbers and three bitwise operations.
The random numbers are denoted by $x_k \quad (k = 1,2,3, \mathrm{and}~4)$.
When the probability is expressed in the binary notation, $0$ corresponds to the bitwise AND ($\land$) and $1$ corresponds to the bitwise OR ($\lor$).
}
\label{fig:finite}
\end{figure}

Suppose $x_k \quad (k =1, 2, \cdots, n)$ is a random bit string such that each bit is set with probability $0.5$ and the target probability $\tilde{p}_n$ is given by,
\begin{equation}
\tilde{p}_n = \left(0.b_n b_{n-1} \cdots b_2 1\right)_\mathrm{bin},
\end{equation}
where $(\cdots)_\mathrm{bin}$ denotes the binary notation.
Note that we should truncate the digits so that the right-most bit (the least significant bit) is one, \textit{i.e.}, $b_1=1$.
We can generate a bit string $\tilde{y}_n$ such that each bit is set with a probability $\tilde{p}_n$
by combining $n$ bit strings as follows.
\begin{align}
\tilde{y}_1 &= x_1, \\
\tilde{y}_{k} &= 
\left\{
\begin{array}{l}
x_k   \lor \tilde{y}_{k-1} \quad \mathrm{if} \quad b_k = 1,\\
x_k  \land \tilde{y}_{k-1} \quad  \mathrm{otherwise}.\\
\end{array}
\right.
\end{align}
The above procedure involves the generation of $n$ random bit strings and $n-1$ bit operations.
See Fig.~\ref{fig:finite} for the schematic illustration of the algorithm.
We refer to this algorithm the finite-digit algorithm.
This algorithm is effective when the number of digits in binary notation of $p$ is small. For example, consider to generate a random bit string which length is $N_\mathrm{bit} = 32$ and in which each bit is set with a probability $p=0.5$.
To generate a such bit string, the binomial-shuffle algorithm involves random number generations $1 + p N_\mathrm{bit} = 17$ times and the Poisson-OR algorithm involves $1-N_\mathrm{bit} \log(1-p) \sim 23$ times in average. However, we can generate a such bit string by calling \verb|std::mt19937| once.

\subsection{Hybrid Algorithm} \label{sec:correction}
\begin{figure}[ht]
\includegraphics[width=7cm]{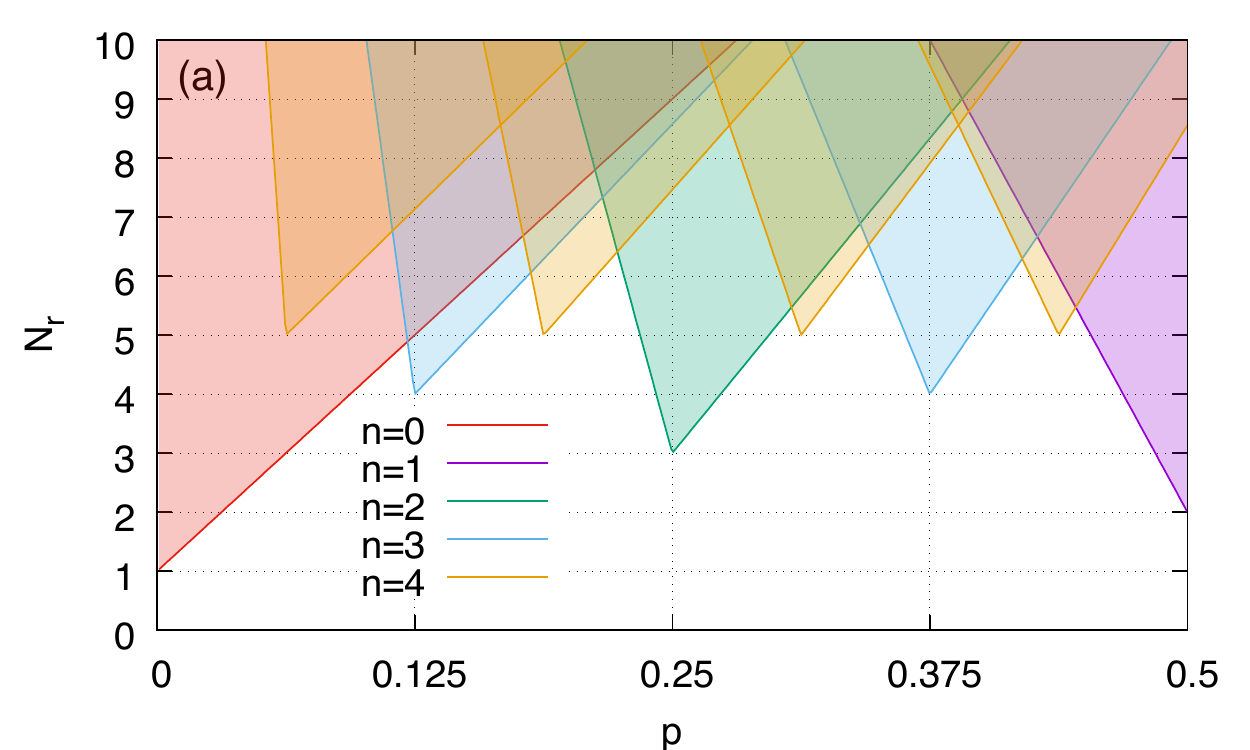}
\includegraphics[width=7cm]{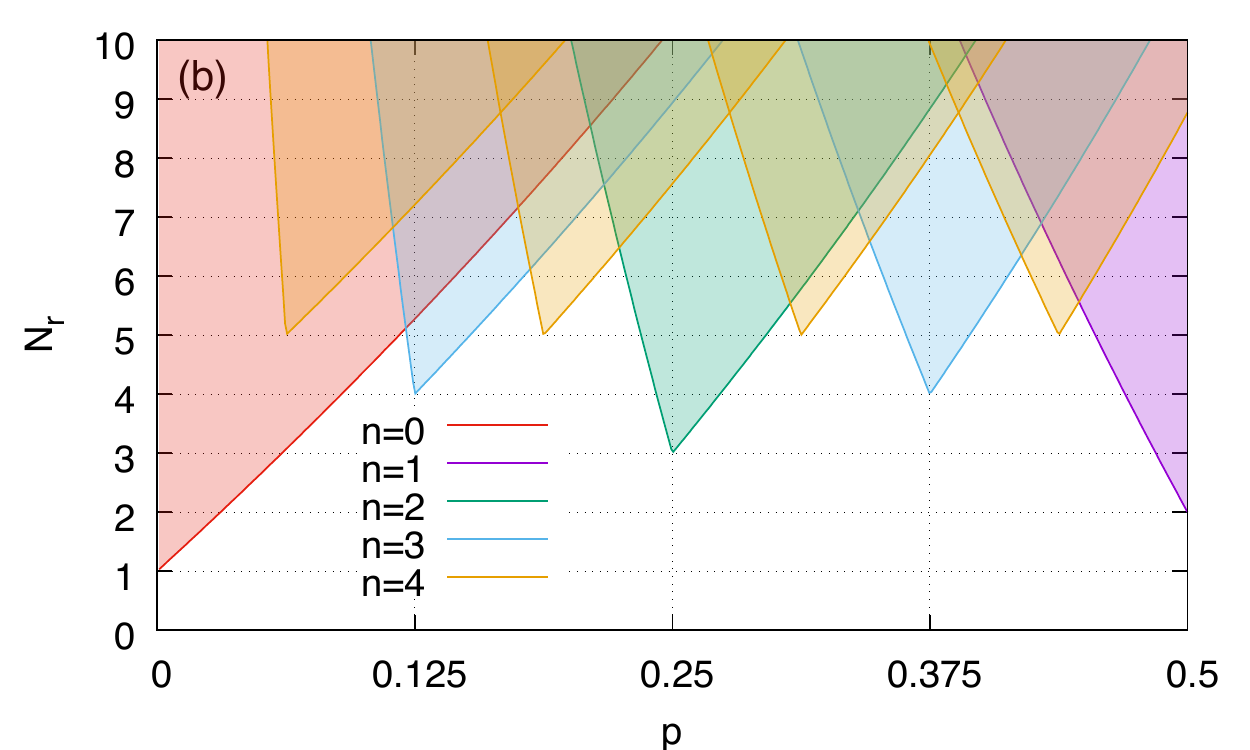}
\caption{
(Color online) 
Expected number of random numbers generated $N_r$
for $n$-digit correction in (a) the binomial-shuffle algorithm
and (b) the Poisson-OR algorithm.
The cases of $N_\mathrm{bit} = 32$ are shown.
}
\label{fig:num}
\end{figure}

The binomial-shuffle and the Poisson-OR algorithms are effective when the number of set bits is small
while the finite-digit algorithm is effective when the number of digits in binary notation of a probability $p$ is small. Combining the two approaches, we can construct a new algorithm which is effective for arbitrary probability. 
As described in the previous section, we can generate a random bit string $\tilde{y}_n$
such that each bit is set with a probability $\tilde{p}_n$.
The bit string $\tilde{y}_n$ can be generated by using $n$ random numbers.
Suppose a target probability $p$ is expressed as $p = \tilde{p}_n + \varepsilon$, where $\varepsilon$ is a small positive number.
We generate the desired bit string $y$ by 
\begin{equation}
y = \tilde{y}_n \lor z,
\end{equation}
where $z$ is a bit string such that each bit is set with probability $p_\varepsilon$.
A bit of $y$ is zero if and only if the bits of $\tilde{y}_n$ and $z$ at the corresponding position are zero. Therefore, the following identity is fulfilled as 
\begin{equation}
1 - p =  (1 - \tilde{p}_n)(1 - p_\varepsilon).
\end{equation}
Solving the above equation with respect to $p_\varepsilon$, we have
\begin{equation}
p_\varepsilon = (p-\tilde{p}_n)/(1-\tilde{p}_n).
\end{equation}

Similarly, we can consider the case $p = \tilde{p}_n - \varepsilon$.
We correct $\tilde{y}_n$ as 
\begin{equation}
y = \tilde{y}_n \land (\lnot z),
\end{equation}
where $\lnot z$ denotes the bitwise NOT of the bit string $z$.
We generate $z$ so that each bit in $z$ is set with the probability $p_\varepsilon$,
therefore, each bit in $\lnot z$ is set with the probability $1-p_\varepsilon$.
A bit of $y$ is set when the bits of $\tilde{y}_n$ and $\lnot z$ at the corresponding position are set.
Therefore, we have
\begin{equation}
p = \tilde{p}_n (1-p_\varepsilon).
\end{equation}
Solving the above with respect to $p_\varepsilon$, we have
\begin{equation}
p_\varepsilon = \frac{\tilde{p}_n - p}{\tilde{p}_n}.
\end{equation}
In both cases, we generate a bit string for the correction, $z$, by the binomial-shuffle or the Poisson-OR algorithms. The probability of the correction $p_\varepsilon$ is $O(|p-\tilde{p}_n|) = O(\varepsilon)$.
Therefore, the number of random numbers that must be generated for the correction becomes small when $\tilde{p}_n$ is close to $p$.

As an example, consider the probability $p=0.6447$, which is the critical point of 1d-BDP~\cite{Jensen1999}, for the case of $N_\mathrm{bit} = 32$.
If we adopt the binomial-shuffle algorithm without the correction, $p N_\mathrm{bit} + 1 \sim 21.6$ random numbers must be generated on average.
The binary notation of $p$ is 
\begin{equation}
p = 0.6447 = (0.101001010\cdots)_\mathrm{bin}.
\end{equation}
First, consider the one-digit correction $p = \tilde{p}_1 + \varepsilon$,
where $\tilde{p}_1 = 0.5$ and $\varepsilon = 0.1447$, respectively.
The probability for the correction is $p_\varepsilon=(0.6447-0.5)/(1-0.5)=0.2894$.
Then the number of random numbers that must be generated to generate a bit string for the correction $z$ is
$p_\varepsilon N_\mathrm{bit} + 1 \sim 10.26$. Therefore, the total number of random numbers generated is $11.26$, which is much smaller than that for the algorithm without correction.

The two-digit correction $p = \tilde{p}_2 + \varepsilon$ is identical to the one-digit correction
since $\tilde{p}_2 = (0.10)_\mathrm{bin} = 1/2 = \tilde{p}_1$.
Therefore, we have to consider the three-digit correction $p = \tilde{p}_3  + \varepsilon = 0.625 + 0.0197$ for the next step.
The bit string $\tilde{y}_3$ is obtained by
\begin{equation}
\tilde{y}_3 = x_3 \lor (x_2 \land x_1).
\end{equation}
Since the probability of the correction is $p_\varepsilon = (0.6447-0.625)/(1-0.625) \sim 0.053$,
the number of random numbers to generarte the bit string for the correction $z$ is $p_\varepsilon N_\mathrm{bit}+1 \sim 2.68$. Since we need three random numbers to generate $\tilde{y}_3$, the total number of random numbers generation is $5.68$.
A finite-digit probability $\tilde{p}_n$ is meaningful when the right-most (least significant) bit is 1. Therefore, the next meaningful step is the six-digit correction which will require at least seven random numbers. Since the three-digits approximation requires $5.68$ random number generations, the three-digits approximation is the most efficient for $p=0.6447$.

We can also consider the correction from the other side, \textit{i.e.}, $p = \tilde{p}_n - \varepsilon$.
The two-digit correction is $p = \tilde{p}_2 - \varepsilon = 0.75 - 0.1053$.
The probability for the correction is $p_\varepsilon = (0.75-0.6447)/0.6447 \sim 0.1633$.
The average number of random numbers generation is $p_\varepsilon N_\mathrm{bit} + 1 + 2 = 8.23$.
Next meaningful step is the four-digit correction $p = 0.6875 - 0.0428$, and 
the average number of random numbers generation is $7.12$.
The next step is the five-digit correction $p = 0.65625 + 0.01155$ which involves $6.57$ random number generations.
Since the next step is the seven-digit correction, the five-digit correction is most effective.
Comparing two approaches, $p = \tilde{p}_n +\varepsilon$ and $p = \tilde{p}_n -\varepsilon$,
the most effective correction is $p = \tilde{p}_3 +\varepsilon$ for $p = 0.6447$.

If the number of digits in the approximation increases, the number of random numbers that must be generated for the correction decreases, whereas that required to generate the initial bit string increases.
Therefore, there is an optimal number of digits for approximation.
Additionally, there are two choices for correction, $p = \tilde{p}_n +\varepsilon$ and $p = \tilde{p}_n -\varepsilon$. The optimal number of digits for the approximation and the expected number of random numbers generated are shown in Fig.~\ref{fig:num}.
This figure shows the case where $0 \le p \le 0.5$. 
One can generate a bit string for $p > 0.5$ by first generating a bit string with probability $1-p$ and 
inverting it.
The expected number of random numbers generated to generate a 32-bit string in which each bit is set with arbitrary probability $p$ is at most $7$. In the case of 64 bits, the expected number of random number generations is at most $8$.

\begin{figure}[ht]
\includegraphics[width=8cm]{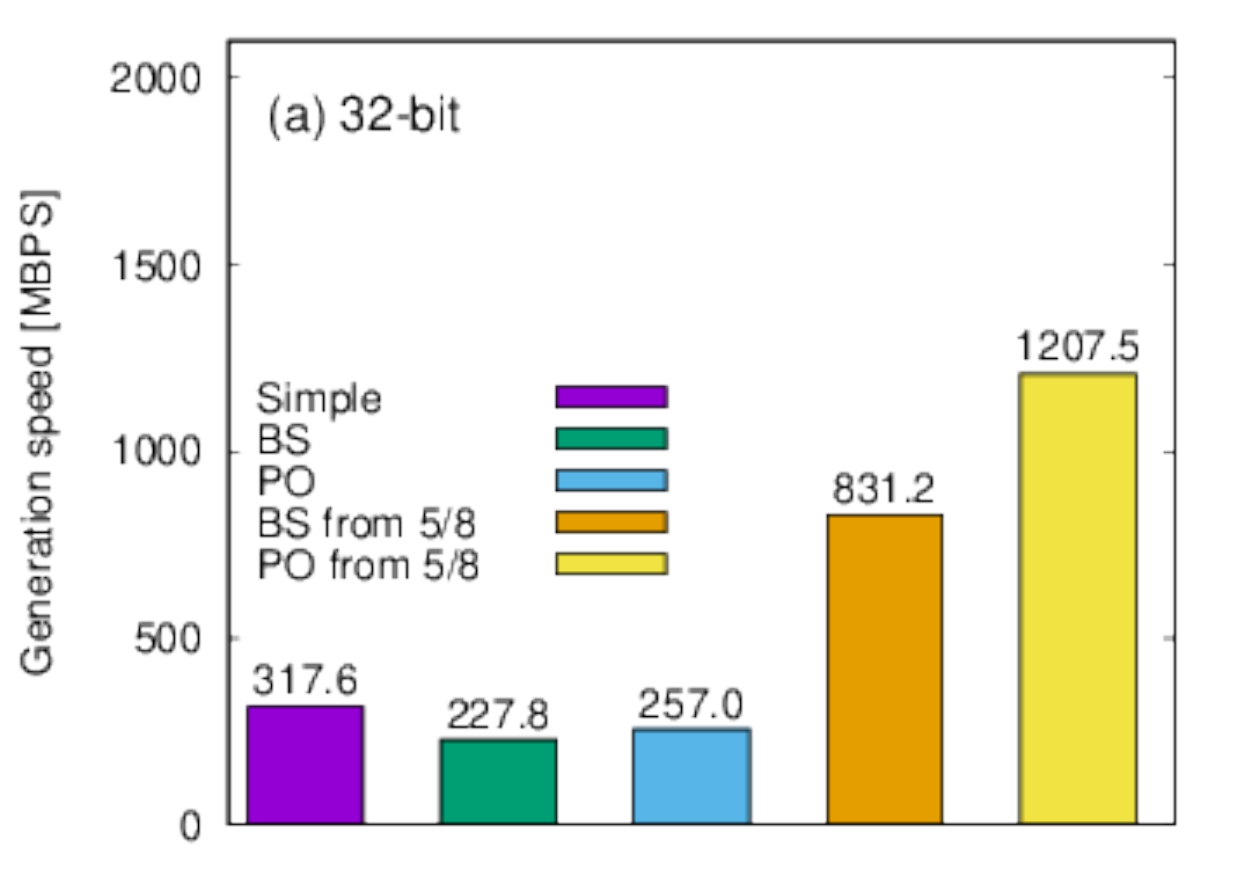}
\includegraphics[width=8cm]{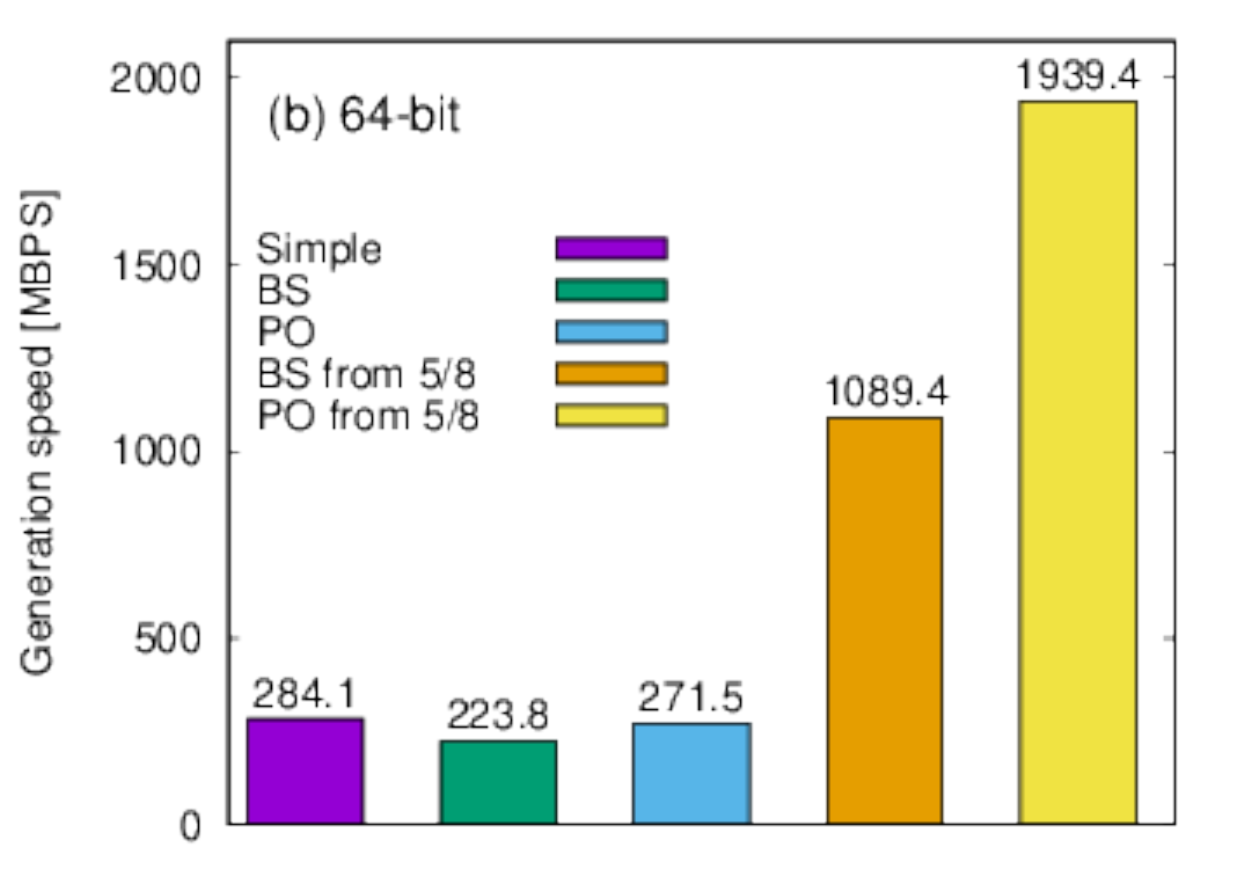}
\caption{
(Color online) Generation speed of random bits set with probability $p=0.6447$ in the unit of MBPS (millions of bits per second) for (a) 32-bit and (b) 64-bit cases.
The results for the simple algorithm (Simple), the binomial-shuffle algorithm (BS), the binomial-shuffle algorithm with the correction from $\tilde{p}_3 = 5/8$ (BS from 5/8), the Poisson-OR algorithm (PO), and the Poisson-OR algorithm with correction (PO from 5/8) are shown.
}
\label{fig:bit}
\end{figure}

\subsection{Benchmark Results} \label{sec:benchmark}

We performed benchmark tests on HPE SGI 8600 system with Intel Xeon Gold 6184 CPU at the Institute for the Solid State Physics of the University of Tokyo.
The program was compiled using Intel C++ compiler 18.0.1 with the option \verb|-O3 -xHOST| and 
executed as a single-threaded process on a single CPU core.
The generation speed of random bits is shown in Fig.~\ref{fig:bit}.
Here, we adopt the unit MBPS (millions of bits per second), which is 1 when one million bits are generated in one second.
When $k$ bit strings of $N_\mathrm{bit}$ width are generated in $t$ [s],
the generation speed is $k N_\mathrm{bit} \cdot 10^{-6} /t$ [MBPS].
We estimate the generation speed by observing the time required to generate $4000000$ bit strings
in which each bit is set with probability $p=0.6447$.
The Poisson-OR algorithm with the correction from the three-digit approximation was the fastest.
Compared with the simple algorithm, the generation speed was about 3.8 times faster for the 32-bit case and 6.8 times for the 64-bit case.
Adopting the 64-bit implementation, the performance of the binomial-shuffle algorithm is improved by 24\%
where that of the Poisson-OR algorithm is improved by 38\%.

\begin{figure}[tb]
\includegraphics[width=8cm]{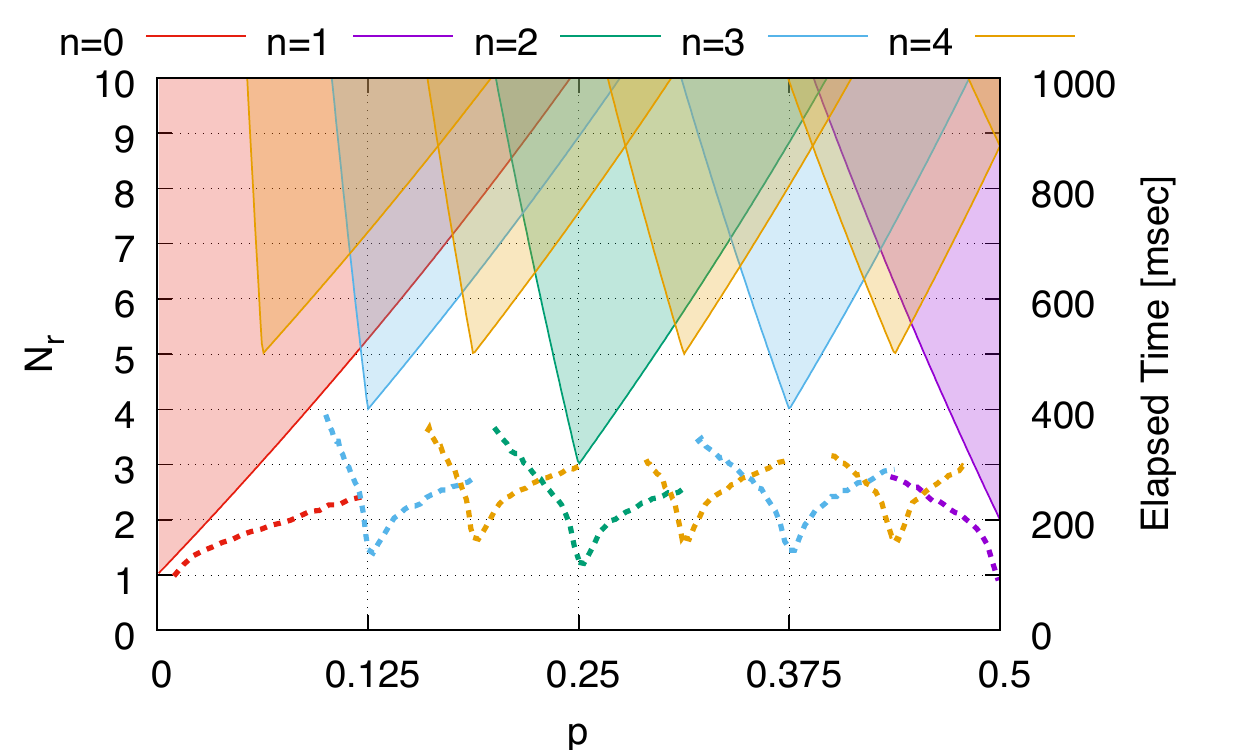}
\caption{
(Color online) Time required to generate random bit strings together with the expected number of random numbers generated $N_r$, which is shown in Fig.~\ref{fig:num}~(b).
The time required to generate 32-bit random bit strings $10^7$ times is denoted by the dashed lines. We adopt the Poisson-OR algorithm with correction. The color corresponds to the number of digits $n$ used in the approximation.
One can see that the performance inversion occurs near the theoretically expected probabilities, \textit{i.e.}, the intersection points of $N_r$.
}
\label{fig:performance}
\end{figure}

The time required to generate a random bit string is expected to be roughly proportional to the number of random numbers generated. When the probability $p$ is close to the $n$-digit-approximated probability $\tilde{p}_n$, then the number of random numbers generated for correction becomes small as shown in Fig.~\ref{fig:num}.
Consider the region $0.125 < p < 0.1875$.
When $p$ is close to $0.125$, then three-digit-approximation from $1/8$ will exhibit the best performance, while
four-digit-approximation from $3/16$ will exhibit the best performance when $p$ is close to $0.1875$.
Therefore, it is expected that the performance inversion will occur between the three- and four-digit approximations in the region $0.125 < p < 0.1875$. 
To demonstrate this, the probability dependence of the time required to generate random bit strings is shown in Fig.~\ref{fig:performance}. 
The performance of the three- and four-digit approximations is reversed at $p=0.1805$, which is close to the expected value of $p=0.181$.

\section{Application to Directed Percolation} \label{sec:dp}

\begin{figure}[tb]
\includegraphics[width=7cm]{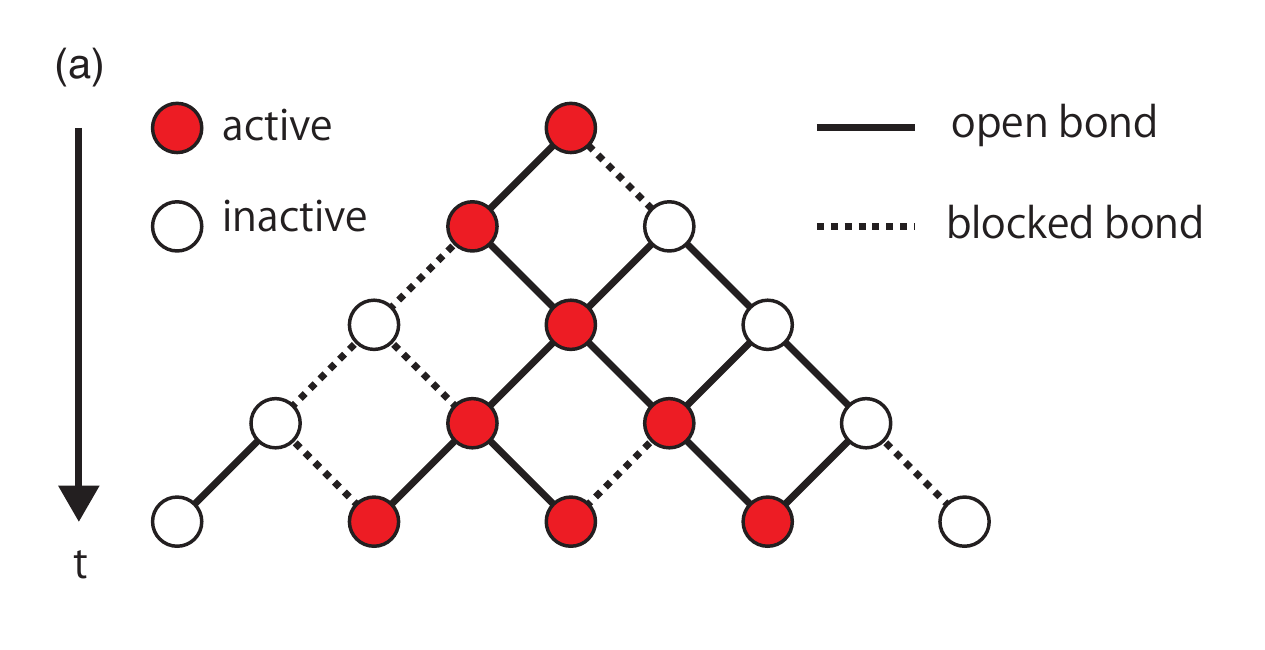}
\includegraphics[width=7cm]{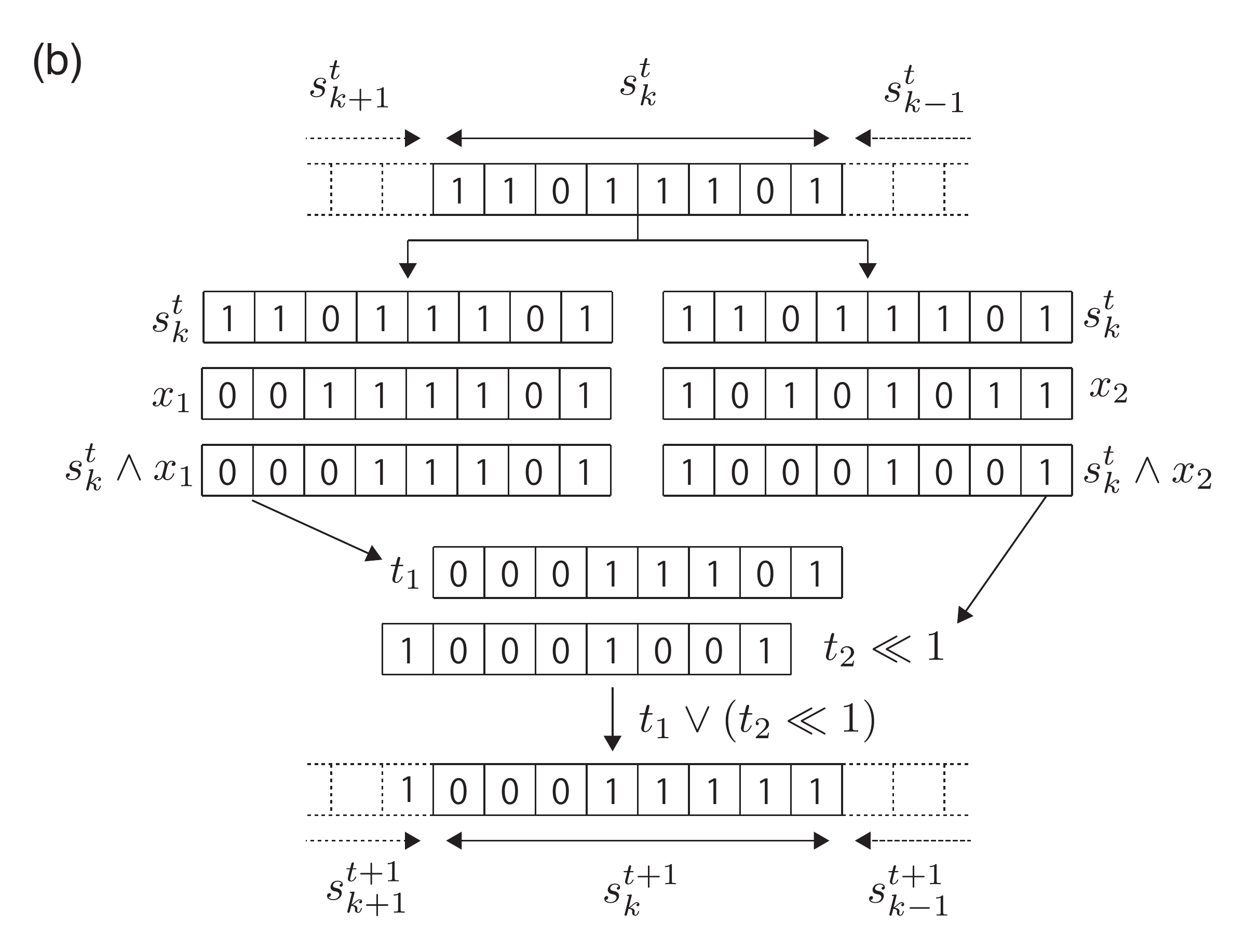}
\caption{
(Color online) (a) Time evolution of 1d-BDP.
Each bond is open (solid lines) with probability $p$ and blocked (dashed lines) otherwise.
If the site on the lower left or lower right of an active site is connected by an open bond, then that site is activated. 
(b)  Implementation of 1d-BDP by bit operations.
Although we performed 32-bit or 64-bit implementation, 8-bit implementation is shown for visibility.
}
\label{fig:dpfig}
\end{figure}

\begin{figure}[tb]
\includegraphics[width=8cm]{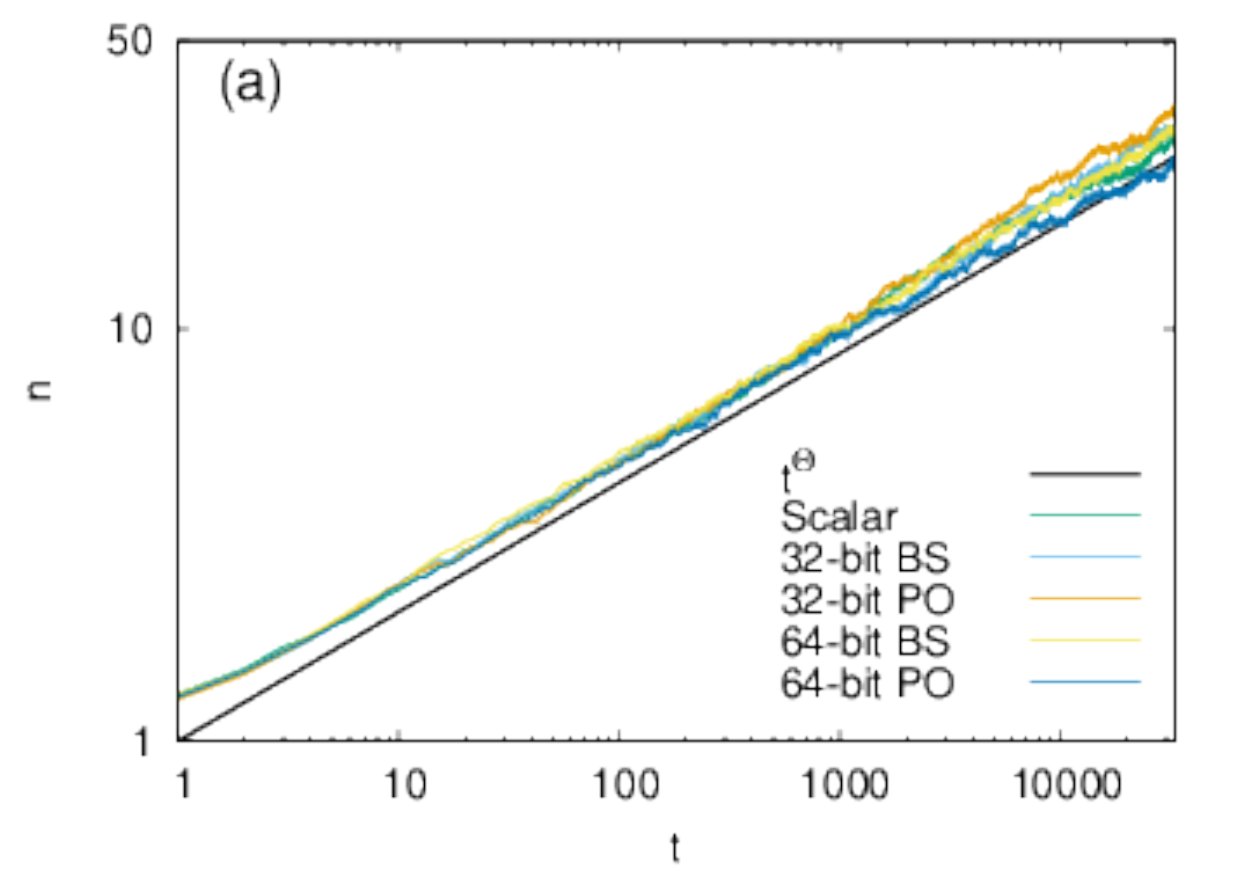}
\includegraphics[width=8cm]{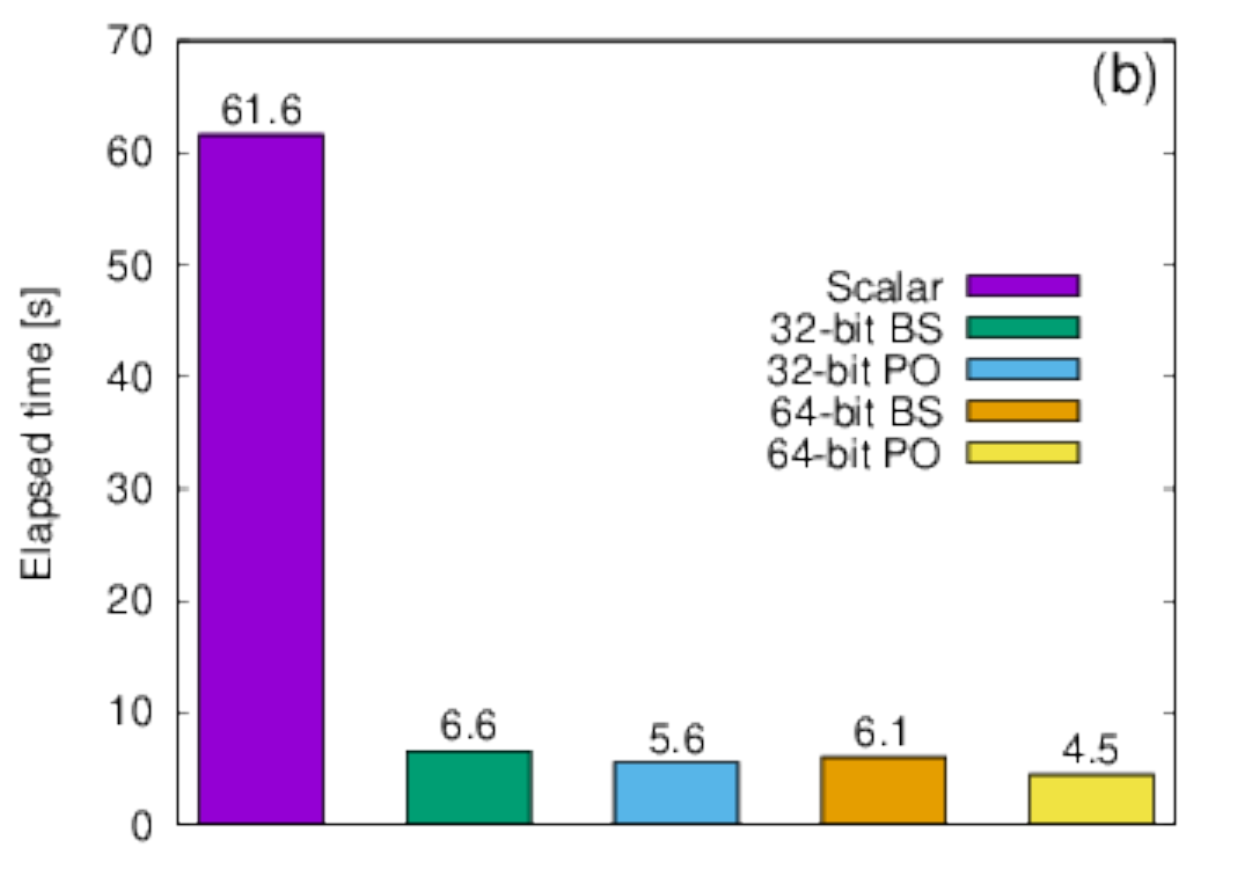}
\caption{
(Color online) 
(a) Cluster growth of 1d-BDP from a single seed at the criticality (log-log axes).
The system size is $L = 32768$.
The time evolution is performed for $32 768$ steps and $10^3$ independent samples are averaged.
The solid line denotes $t^\Theta$, where $\Theta=0.313$.
The results with the scalar code (Scalar) and bit-operation implementation with the binomial-shuffle algorithm (BS) and Poisson-OR algorithm (PO) are shown.
We adopt the correction from the three-digit approximation $\tilde{p_3} = 5/8$ for both the BS and PO algorithms.
(b) Time required for the simulations.
}
\label{fig:cluster}
\end{figure}

\begin{figure}[tb]
\includegraphics[width=8cm]{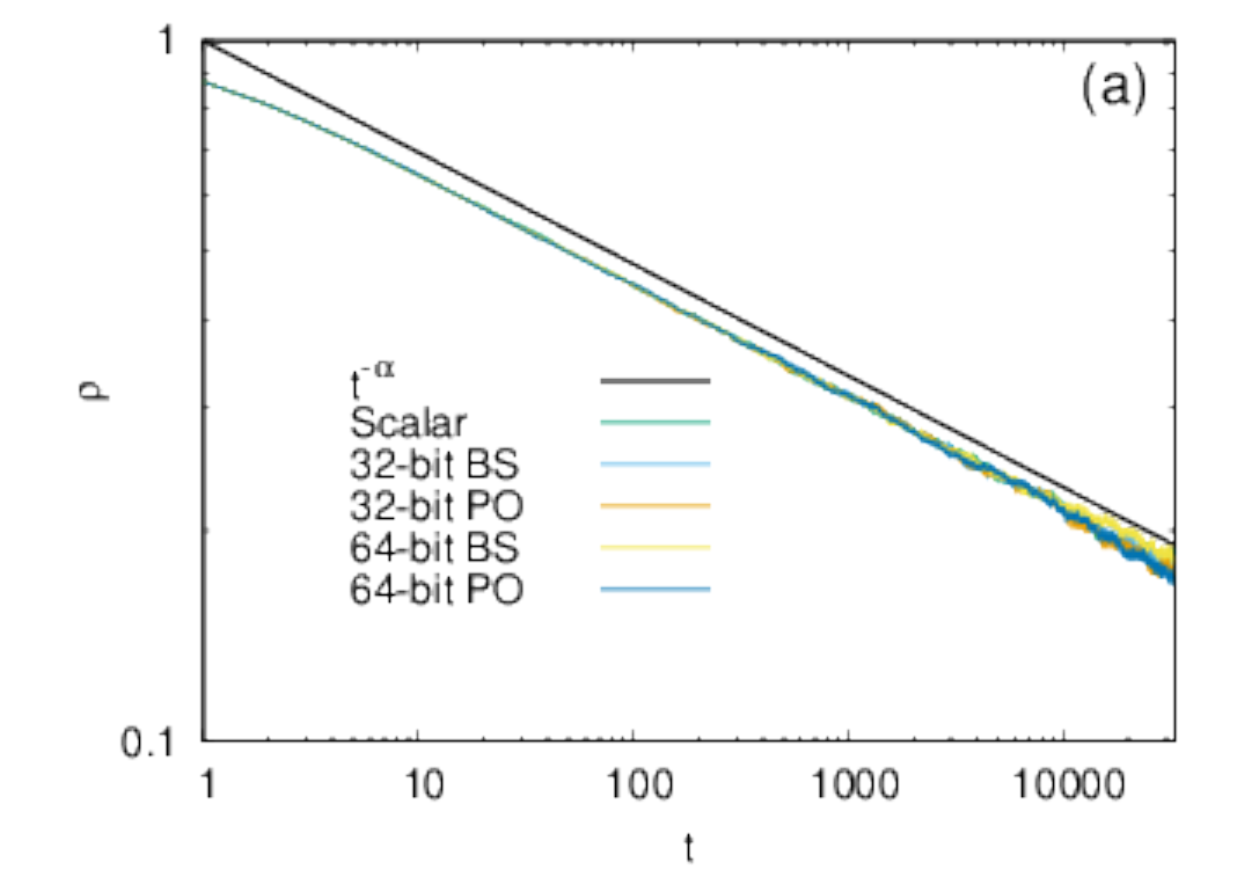}
\includegraphics[width=8cm]{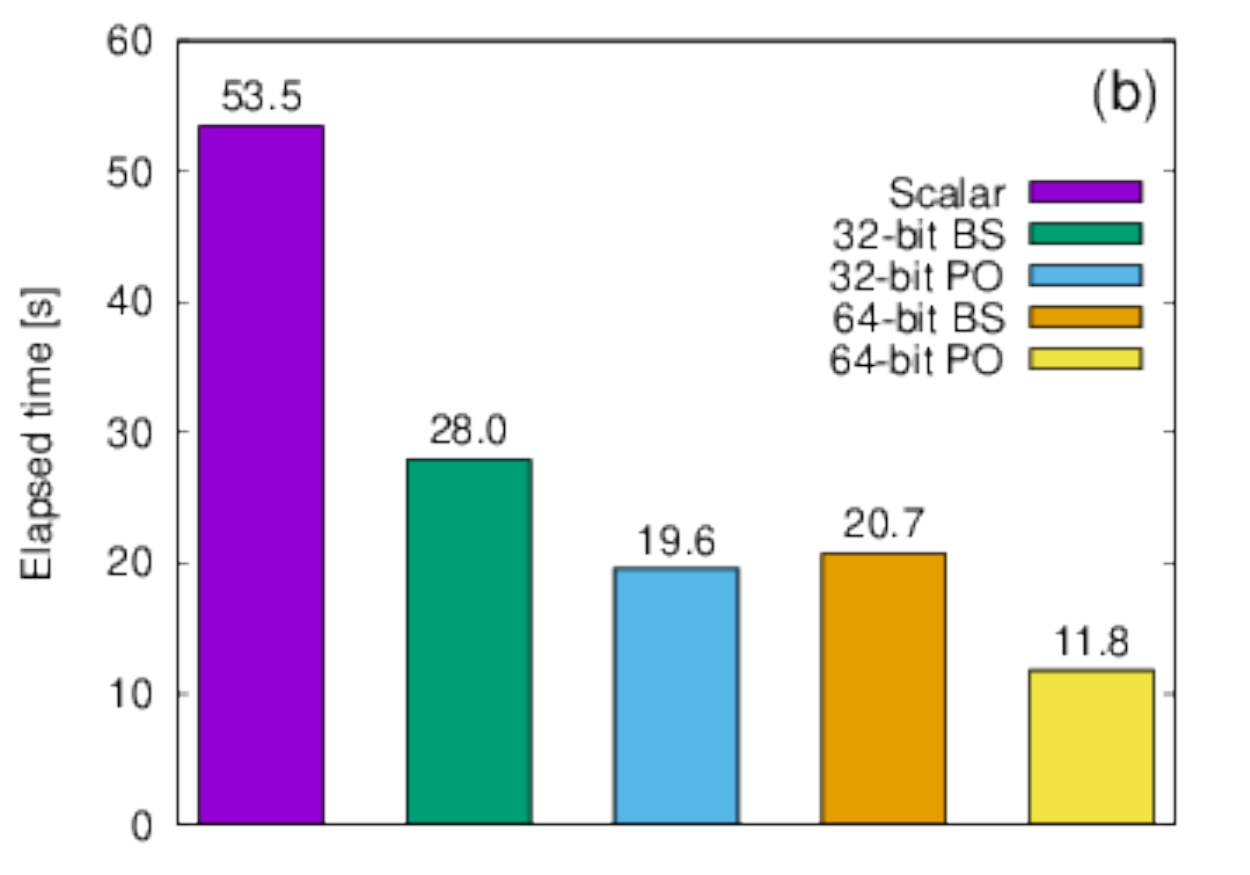}
\caption{
(Color online) 
(a) Decay of the density of 1d-BDP at the criticality (log-log axes).
The time evolutions of the densities of the active sites from the fully active state are shown.
The system size is $L = 32768$.
The time evolution is performed for $32768$ steps and $10$ independent samples are averaged.
The solid line denotes $t^{-\alpha}$ where $\alpha=0.159$.
The results with the scalar code (Scalar) and bit-operation implementation with the binomial-shuffle (BS), the Poisson-OR (PO) algorithms are shown. We adopt the correction from the three-digit approximation $\tilde{p_3} = 5/8$ for both BS and PO algorithms.
(b) Time required for the simulations.
}
\label{fig:periodic}
\end{figure}

Employing the fast random bit-string generation algorithms developed in the previous section,
we apply MSC to 1d-BDP.
The time evolution of 1d-BDP is shown in Fig.~\ref{fig:dpfig}~(a).
Each bond is open with probability $p$ and blocked otherwise.
The state of the $i$th site at time $t$ is denoted by $\sigma_i^t$.
The site is active when $\sigma_i^t=1$ and inactive when $\sigma_i^t=0$.
If a site at time $t+1$ is connected to an active site at time $t$ with an open bond, then that site becomes active. The states of the sites at time $t+1$ are determined by the states at time $t$
and the time evolution is performed by iterating this process.
The scalar implementation to determine the states of the sites at time $t+1$ from the states at time $t$ is shown in Algorithm~\ref{alg:dp_scalar}. Here, we omit the processing of the boundary conditions.

\begin{algorithm}
\begin{algorithmic}[1]
\For{ $i=1$ to $L$}
\If {$\sigma_i^t= 1 $} 
\If {$U_r(0,1) <p$} 
\State $\sigma_i^{t+1}= 1$
\EndIf
\If {$U_r(0,1) <p$} 
\State $\sigma_{i+1}^{t+1}= 1$
\EndIf
\EndIf
\EndFor
\end{algorithmic}
\caption{Scalar Implementation of 1d-BDP}
\label{alg:dp_scalar}
\end{algorithm} 

We pack the information of $N_\mathrm{bit}$ sites in a bit string, where $N_\mathrm{bit} =32$ or $64$.
If a bit is set, then the corresponding site is active and inactive otherwise.
The $k$th bit string at time $t$ is denoted by $s_k^t$.
The bit string $s_k^t$ contains the information of $\{\sigma_i^t\}$, where $i=(k-1)N_\mathrm{bit}+1, \cdots, kN_\mathrm{bit}$.
We generate a bit string $x_1$ in which each bit is set with probability $p$
and take the bitwise AND between $s_k^t$ and $x_1$ as $t_1=s_k^t \land x_1$.
Then each bit of $s_k^t$ survives to $t_1$ with probability $p$.
Therefore, $t_1$ can be considered as the active sites at time $t+1$ 
connected to the active sites at time $t$ with the lower left bond.
Similarly, we generate $t_2 =s_k^t \land x_2$, which denotes 
the active sites at time $t+1$ connected to the active sites at time $t$ with the lower right bond.
Then the site configuration at time $t+1$ is obtained by taking the 
bitwise OR between $t_1$ and $(t_2 \ll 1)$.
Note that the most significant bit of $t_2$ should be copied to the least significant bit of $s^{t+1}_{k+1}$.
A schematic illustration of the implementation is shown in Fig.~\ref{fig:dpfig}~(b)
and its pseudocode is shown in Algorithm~\ref{alg:dp_bit}.
The operation $a \gg k$ denotes the logical (zero-fill) right shift which divides $a$ by $2^k$ with rounding towards zero.
$\mathrm{RBS}(p)$ denotes a random bit string of length $N_\mathrm{bit}$ in which 
each bit is set with probability $p$.
The processing of the boundary conditions is omitted.

\begin{algorithm}
\begin{algorithmic}[1]
\For{$k=1$ to $L/N_\mathrm{bit}$}
\State $x_1 \leftarrow \mathrm{RBS}(p)$
\State $x_2 \leftarrow \mathrm{RBS}(p)$
\State $t_1 \leftarrow s_k^t \land x_1$
\State $t_2 \leftarrow s_k^t \land x_2$
\State $s_k^{t+1} \leftarrow s_k^{t+1} \lor t_1$
\State $s_k^{t+1} \leftarrow s_k^{t+1} \lor (t_2 \ll 1)$
\State $s_{k+1}^{t+1} \leftarrow (t_2 \gg (N_\mathrm{bit}-1))$
\EndFor
\end{algorithmic}
\caption{MSC Implementation of 1d-BDP}
\label{alg:dp_bit}
\end{algorithm} 

We perform the benchmark simulations of 1d-BDP using the MSC implementation.
The benchmark conditions are the same as those in Sec.~\ref{sec:benchmark}.
The cluster growth from a single seed at the criticality $p=0.6447$ is shown in Fig.~\ref{fig:cluster}~(a).
The system size is $L= 32768$, the time evolution is performed for $32768$ steps and $10^3$ independent samples are averaged. 
Power-law behavior is observed for the number of active sites $n(t) \sim t^\Theta$, where $\Theta=0.313$~\cite{Henkel2009}.
The time required to perform these simulations is shown in Fig.~\ref{fig:cluster}~(b).
The fastest algorithm was the Poisson-OR algorithm with the correction, whose speed was 14 times that of the scalar algorithm.

The relaxation process from the all-active state at the criticality $p=0.6447$ is shown in Fig.~\ref{fig:periodic}~(a). 
The system size is $L= 32 768$ with the periodic boundary condition.
The time evolution is performed for $32 768$ steps and $10$ independent samples are averaged.
Power-law decay $t^{-\alpha}$ is observed for the density of the active sites $\rho(t) = n(t)/L$, where $\alpha =0.159$~\cite{Henkel2009}.
The time required to perform these simulations is shown in Fig.~\ref{fig:periodic}~(b).
While the fastest algorithm is the Poisson-OR algorithm with the correction, in this case, 
the increase in speed is only 4.5 times, which is much smaller than that in the cluster growth simulation.

\section{Summary and Discussion} \label{sec:summary}

We have developed efficient algorithms to generate a bit string in which each bit is set with arbitrary probability. The binomial-shuffle algorithm first determines the number of set bits and then determines their position. The Poisson-OR algorithm first determines the number of bit strings in which one bit is set randomly and then takes the bitwise OR between them. While the expected number of generated random numbers is smaller in the binomial-shuffle algorithm than in the Poisson-OR algorithm, the Poisson-OR algorithm is faster owing to the simple loop structure. The finite-digit algorithm allows us to generate a random bit string with a finite-digit probability in the binary notation.
Combining two algorithms, the number of the random numbers that must be generated for the correction markedly decreases.

We developed the MSC technique for 1d-BDP using the random bit string generation algorithms and achieved a marked increase in speed.
The MSC was more effective in the simulation of cluster growth from a single seed than in that of the relaxation from the fully active state. This is due to the local density of the active sites.
The number of random numbers generated for the scalar algorithm is proportional to the density of active sites, while that for MSC is independent of the density. Therefore, the efficiency of MSC decreases as the density of active sites decreases. Since the density of active sites in the relaxation process decreases monotonically, the efficiency of MSC decreases over time. In the case of cluster growth simulation, we only update the region between the leftmost to rightmost active sites. Then the local density of active sites hardly changes and MSC works effectively for this case.
Our random bit generating algorithm can be applied to replica-parallel MSC.
While the previous implementation of MSC for DP used same random numbers among replicas, the replicas do not share the random numbers with our algorithm.

The random bit string generation algorithm is expected to be applicable to general Monte Carlo simulations on lattice systems.
In the present work, we did not consider the use of SIMD instructions.
SIMD stands for single instruction multiple data and it allows data-level parallelism with the SIMD register. For example, a 512-bit register is available in the Intel Advanced Vector Extensions (AVX-512). By using SIMD instructions, the efficiency of MSC can be further improved. 
In recent years, general-purpose computing on graphics processing units (GPGPU) has attracted the interests of many researchers. The MSC of the Ising model was implemented on GPGPU~\cite{Block2010}, and Komura and Okabe implemented the Swendsen--Wang algorithm on GPGPU~\cite{Komura2012}. The implementation of the MSC algorithm presented in the manuscript on GPGPU should also be attempted in the future.

\section*{Acknowledgements}
The authors would like to thank K. Harada, T. Suzuki, T. Okubo, and R. Kaneko for helpful discussions.
This work was supported by JSPS KAKENHI Grant Number 15K05201 and by the MEXT project ``Exploratory Challenge on Post-K Computer'' (Frontiers of Basic Science: Challenging the Limits). The computations were carried out using the facilities of the Institute for Solid State Physics of the University of Tokyo.

\section*{Appendix}

In this appendix, we show the equivalence of the Poisson-OR algorithm
(Algorithm \ref{alg:or}) and the simple algorithm (Algorithm
\ref{alg:simple}).  In other words, we prove that the probability of
observing a particular $N$-length bit string $s$ in the Poisson-OR algorithm is given by
\begin{equation}
 P(s) = p^{m} (1-p)^{N-m}, \label{eq:p_simple}
\end{equation}
where $m$ is the number of set bits in $s$.  This fact indicates that
each bit is set with probability $p$ mutually independently.

First, let us consider the number of ways to classify $k$ labeled
elements to $m$ unlabeled groups so that no empty group exists.  This
number is denoted by $S(k, m)$ which is called the Stirling numbers of
the second kind.  The explicit expression of $S(k, m)$ is given by,
\begin{equation}
S(k,m) = \frac{1}{m!} \sum_{j=0}^m (-1)^{m-j} {m\choose j} j^k, \label{eq:s_general}
\end{equation}
where ${m\choose i}$ is a binomial coefficient.  We define $S(k,m) = 0$
when $k<m$.  It is useful to derive the exponential generating function
of $S(k,m)$.  According to the binomial theorem, the following identity
holds.
\begin{equation}
(\mathrm{e}^x - 1)^m = \sum_{j=0}^m (-1)^{m-j}  {m\choose j}\mathrm{e}^{jx}.
\end{equation}
Differentiating both sides $k$ times with respect to $x$ and then
substituting $x = 0$, we have
\begin{equation}
\begin{split}
\left. \left(\frac{d}{dx} \right)^k (\mathrm{e}^x - 1)^m \right|_{x=0} &=
\sum_{j=0}^m (-1)^{m-j}  {m\choose j} j^k \\
&= m! S(k,m).
\end{split}
\end{equation}
Therefore, we have the exponential generating function of the Stirling
numbers as
\begin{equation}
 \sum_{k=0}^\infty S(k,m) \frac{x^k}{k!}
  =\frac{1}{m!} (\mathrm{e}^x - 1)^m. \label{eq:egf}
\end{equation}

Next, consider $N$-length bit strings each of which has a single set bit
randomly, and suppose that bitwise OR between $k$ bit strings yields a
bit string $s$ which has $m$ set bits.  Since the position of set bits
in $s$ is labeled, the number of possible configurations of bit strings
is $m! S(k,m)$.  Thus the probability $P_k(s)$ of obtaining a bit string
$s$ is given by
\begin{equation}
P_k(s) = \frac{m! S(k,m)}{N^k}.
\end{equation}

In the Poisson-OR algorithm, we choose the number of bit strings to be
taken bitwise OR following the Poisson distribution with the parameter
$\lambda$. Then the probability that the number of bit strings becomes
$k$ is $\lambda^k \mathrm{e}^{-\lambda}/k!$.  Therefore, the probability
$P(s)$ that the resulting bit string becomes $s$ is
\begin{equation}
\begin{split}
P(s) &= \sum_{k=0}^\infty P_k(s) \frac{\lambda^k \mathrm{e}^{-\lambda}}{k!}, \\
 &= \mathrm{e}^{-\lambda} m! \sum_{k=0}^\infty
 S(k,m) \frac{(\lambda/N)^k}{k!}.  \label{eq:pk}
\end{split}
\end{equation}
From Eq.~(\ref{eq:egf}), we have
\begin{align}
P(s) &= \mathrm{e}^{-\lambda} (\mathrm{e}^{\lambda/N}-1)^m.
\end{align}
Since the parameter of the Poisson distribution is $\lambda = - N
\log(1-p)$, we finally obtain Eq.~(\ref{eq:p_simple}).

\bibliography{dpbit}

\end{document}